\documentclass[journal=cmatex,manuscript=article]{achemso}

\usepackage{chemformula} 
\usepackage[T1]{fontenc} 
\usepackage{textgreek} 
\usepackage[font=bf]{caption} 	
\usepackage{lipsum}
\usepackage{multirow}
\SectionNumbersOn 
\setkeys{acs}{keywords = false}
\setkeys{acs}{articletitle = false}

\author{Wang Lu} \affiliation[National University of Singapore]
{Department of Materials Science and Engineering, National University of
Singapore, 9 Engineering Drive 1, 117575, Singapore}

\author{Juefan Wang}\affiliation[National University of Singapore]
{Department of Materials Science and Engineering, National University of
Singapore, 9 Engineering Drive 1, 117575, Singapore}

\author{Gopalakrishnan Sai Gautam} \affiliation[Indian Institute of Science]
{Department of Materials Engineering, Indian Institute of Science, Bangalore 560012, India}
\email{saigautamg@iisc.ac.in}

\author{Pieremanuele Canepa}
\affiliation[National University of Singapore]
{Department of Materials Science and Engineering, National University of
Singapore, 9 Engineering Drive 1, 117575, Singapore}
\altaffiliation{Department of Chemical and Biomolecular Engineering, National University of Singapore, 4 Engineering Drive 4, 117585, Singapore}
\email{pcanepa@nus.edu.sg}

\title{Searching Ternary Oxides and Chalcogenides as Positive Electrodes for Calcium Batteries}

\begin{document}




\begin{abstract} 
\noindent The identification of  alternatives to the Lithium-ion battery architecture remains a crucial priority in the diversification of energy storage technologies. 
Accompanied by the low reduction potential of \ch{Ca^{2+}/Ca}, --2.87~V vs.\ SHE, metal-anode-based rechargeable Calcium (Ca) batteries appear competitive in terms of energy densities. However, the development of Ca-batteries lacks high-energy density intercalation cathode materials. Using first-principles methodologies, we screen a large chemical space for potential Ca-based cathode chemistries, with composition of \ch{Ca_iTM_jZ_k}, where  \ch{TM} is a  1$^{st}$ or  2$^{nd}$  row transition metal and \ch{Z} is oxygen, sulfur, selenium or tellurium. 10 materials are selected and  their Ca intercalation properties are investigated. We identify two previously unreported promising electrode compositions: the post-spinel \ch{CaV2O4}  and  the layered \ch{CaNb2O4}, with Ca migration barriers of $\sim$654~meV and $\sim$785~meV, respectively. Finally, we analyse the geometrical features of the Ca migration pathways across the 10 materials studied and provide an updated set of design rules for the identification of good ionic conductors, especially with large mobile cations. 
 \end{abstract}

\section{Introduction}

The lithium (Li)-ion battery technology is well-established and has been commercialized for the past three decades, with important but incremental improvements in terms of its energy density, reliability, and safety.\cite{Tarascon2010,Noorden2014,Larcher2015,Nykvist2015,Cano2018,Crabtree2019,Tian2020} 
However, the energy density prerequisites to drive stationary energy storage devices and electric vehicles  are quickly exceeding the limits of what can be achieved by commercial Li-ion batteries.\cite{Whittingham2014} Finding suitable alternatives to the rocking-chair Li-ion battery is thus a crucial priority in the diversification and modernisation of energy storage technologies.  Rechargeable energy-storage devices based on multivalent ions, e.g., magnesium \ch{Mg^{2+}}, calcium \ch{Ca^{2+}}, and zinc \ch{Zn^{2+}}, have emerged as interesting alternatives to their \ch{Li}-ion analogues.\cite{Muldoon2014,Canepa2017} In particular, metal-anode-based rechargeable \ch{Ca} batteries appear competitive in terms of energy densities, given the high gravimetric capacity, $\sim$2205~mAh/g, low standard reduction potential of \ch{Ca^{2+}/Ca}, --2.87~V vs.\ SHE, and large volumetric capacity, $\sim$2073~Ah/l.\cite{Muldoon2014,Canepa2017,Dompablo2019,Kharbachi2020,Blanc2020} 

While a Ca-based chemistry is promising in theory, a number of unresolved challenges stand out. First of all, the lack of practical electrolytes that can simultaneously tolerate the reducing environment of a Ca-metal anode and the oxidizing nature of a high-voltage (and high-capacity) Ca-intercalation cathode.\cite{Peled1981,Aurbach1991,See2013,Ponrouch2015,Wang2017,Hahn2018,Li2019,Shyamsunder2019,Scafuri2020,Bitenc2020,Yang2020,Gao2020,Liang2020,ForeroSaboya2020,Hahn2020,Hahn2020a,Li2020,Xu2021} Another significant challenge is the identification of fast \ch{Ca^{2+}} conductors that can act as reversible, high-voltage, and high-capacity intercalation cathodes, which is the focus of this work.\cite{Rong2015,Chen2019,Boelle2020} 

A reasonable figure of merit to identify potential Ca-cathodes with high Ca-conductivity is represented by the migration barrier, $E_m$ ---the energy that \ch{Ca^{2+}} must overcome during its migration between two identical crystallographic sites in an anion host framework.  Rong et al.\cite{Rong2015} proposed values of $\sim$525--650~meV as tolerable limits for $E_m$ in electrodes, where the lower (upper) limit represents micron (nano) sized electrode particles operating at room temperature and a C/2 rate. If the cathode particle sizes are decreased or the temperature of operation is increased or the rate of battery (dis)charge decreased, the upper limit on $E_m$ can be further increased.\cite{Chen2019} For example, the upper limit on $E_m$ can be increased up to $\sim$895~meV if the cathode particles of size 10~nm are cycled at 333~K ($\sim$60 $^{\circ}$C) at a C/6 rate (6 hours of charge/discharge).\cite{Chen2019}  

Implementing the $E_m$ criteria, Rong et al.\cite{Rong2015} investigated a number of close-packed frameworks predominantly used in Li-ion batteries, i.e., the spinel (Mn$_2$O$_4$), the layered (NiO$_2$), and the olivine (FePO$_4$) prototypes, respectively. Three design rules to facilitate the identification of fast multivalent-conductors were discovered, based on the cation coordination environments and the volume per anion available in a host framework (see Sec.~\ref{sec:discussion} for a detailed description).\cite{Rong2015,Canepa2017a}  Given the large values ($> 525 $~meV) of $E_m$ estimated for Ca migration in layered and olivine frameworks, and the intrinsic instability of Ca-containing spinel frameworks, the authors concluded that these structures are unsuitable for Ca batteries.\cite{Rong2015}  

Recently, Vaughey and coworkers reported reversible intercalation of 0.6 mol \ch{Ca^{2+}} into  NaSICON-\ch{NaV2(PO4)3}, equivalent to a capacity of $\sim$81~mAh/g at 3.2 V vs.\ \ch{Ca}/\ch{Ca^{2+}},\cite{Kim2020} while similar experiments with olivine-\ch{FePO4} yielded partially reversible \ch{Ca} intercalation, with an initial capacity of $\sim$72~mAh/g at $\sim$2.9~V vs.\ \ch{Ca}/\ch{Ca^{2+}}.\cite{Kim2020} These experimental results on olivine-\ch{FePO4} are in qualitative agreement with Rong et al.,{\cite{Rong2015}} who predicted a moderate Ca migration barrier ($\sim$580 meV) in the charged state of olivine-{\ch{FePO4}}. Recently, the group of Kang could reversibly extract  \ch{Ca^{2+}}-ions  from \ch{Na_{1.5}VPO_{4.8}F_{0.7}}  and reported a capacity of 87~mAh~g$^{-1}$ and 90\% capacity retention over 500 cycles.\cite{Xu2021} 

Additional theoretical work\cite{Rong2015,Gautam2015} revealed that $E_m$ for Ca in $\delta$-\ch{V2O5} could be as low as $\sim$200~meV. However, $\delta$-\ch{V2O5} is not the stable polymorph, with $E_m$ for Ca being predicted to be prohibitively high ($\sim$1700--1900~meV) in the stable $\alpha$-\ch{V2O5} polymorph.\cite{Gautam2015}  Although Ca intercalation in $\alpha$-{\ch{V2O5}} has received significant attention,{\cite{Amatucci2001,Verrelli2018,Murata2019}} Verrelli et al.{\cite{Verrelli2018}} confirmed the unsuitability of {\ch{V2O5}} as a cathode for Ca batteries owing to lack of robust Ca intercalation, chemically or electrochemically.

Nevertheless, the search for potential Ca cathodes has progressed steadily in the recent past with a handful of research groups fully dedicated to the task.  For example, the teams led by Palac\'{i}n and Ponrouch have extensively investigated \ch{TiS_2} using different temperatures, electrolyte formulations, and observed reversible Ca intercalation (accompanied by solvent co-intercalation) from a  \ch{Ca(TFSI)2}:PC electrolyte at moderate temperatures ($\sim$60~$^{\circ}$C and 100~$^{\circ}$C).\cite{Tchitchekova2018,Verrelli2020} The same teams have also reported the irreversible Ca extraction from \ch{Ca_{0.89}TaN2}.\cite{Verrelli2019} Tirado and collaborators  observed Ca intercalation from a non-aqueous electrolyte into the hydrated phase of layered-$\alpha$-\ch{MoO3},\cite{Cabello2018} and claimed a gravimetric capacity of $\sim$100~mAh/g at a low voltage of $\sim$1.3~V vs.\ \ch{Ca}/\ch{Ca^{2+}}, with additional corroboration from computational models.

On the computational front, using density functional theory (DFT{\cite{hohenberg_inhomogeneous_1964, Kohn1965}}) Arroyo-de Dompablo et al.\ spearheaded the investigation (and discovery) of alternative Ca-intercalation chemistries, including oxide (i.e.,  \ch{CaMn2O4}, \ch{Ca2Mn2O5}, \ch{CaMn4O8}, \ch{Ca2Fe2O5}, {\ch{Ca4Fe9O17}}, {\ch{Ca3Co4O9}} and {\ch{Ca3Co2O6}}) and mixed anion (i.e., \ch{CaFeSO}, \ch{CaCoSO}, \ch{CaNiN}, \ch{Ca3MnN3}, \ch{Ca2Fe(Si2O7)}, \ch{CaM(P2O7)} with \ch{M = V}, \ch{Cr}, \ch{Mn} \ch{Fe} and \ch{Co}, \ch{CaV2(P2O7)2}, \ch{Ca(VO)2(PO4)2}, and $\alpha$-\ch{VOPO4}) frameworks.\cite{Dompablo2016,Dompablo2016a,Torres2019,Torres2021,Torres2019a,Black2020}  For example, they reported a low Ca $E_m$ of $\sim$650~meV in $\alpha$-\ch{VOPO4}, along with the high theoretical gravimetric capacity ($\sim$312~mAh/g) and average (theoretical) intercalation voltage of $\sim$2.8~V vs.\ \ch{Ca}/\ch{Ca^{2+}}, making \ch{VOPO4} a promising candidate for further experimental investigation. 

Undoubtedly, the above experimental and theoretical studies have made sizeable progress towards the development of practical Ca-battery prototypes. However, a more streamlined theoretical (and experimental) effort can accelerate the search for cathodes with facile \ch{Ca^{2+}} migration, thus facilitating the development of this promising energy storage platform.  Here, we have used an efficient screening of Ca-containing ternary structures, utilising a combination of structural information, characterization conditions, and theoretically-predicted thermodynamic stability, and subsequently evaluated the battery properties of key candidates using DFT-based calculations to push forward the discovery of novel Ca-cathodes. We focus on unveiling overlooked potential cathode chemistries with a general composition of \ch{Ca_iTM_jZ_k}, where  \ch{TM} is a  1$^{st}$ or  2$^{nd}$  row transition metal and \ch{Z} is oxygen, sulfur or selenium. In the present work, we have not considered any low gravimetric-capacity, mixed anion frameworks or any halogen-based structures. 

From our search, we propose two new promising candidates, namely post-spinel \ch{CaV2O4} and the layered \ch{CaNb2O4}.  Specifically, the calculated $E_m$ in both \ch{CaV2O4} ($\sim$654~meV) and  \ch{CaNb2O4}  ($\sim$785~meV) fulfil the criteria for Ca-conductors with reasonable Ca-conductivity. As further validation of our proposed screening strategy, we re-discovered the Chevrel-\ch{Mo6S8} and \ch{CaMoO3} compounds, which have already been studied both computationally and experimentally.\cite{Rogosic2014,Dompablo2016a,Cabello2018} Finally, we revisit and propose modifications to the design rules proposed by Rong \emph{et al.}  for good multivalent conductors,\cite{Rong2015}  which can be further extended to discover novel conductors for Ca-batteries and beyond.

\section{Results}

\subsection{Screening Procedure of \ch{Ca_iTM_jZ_k} Electrode  Materials}

\begin{figure*}[ht!] 
\includegraphics[width=\textwidth]{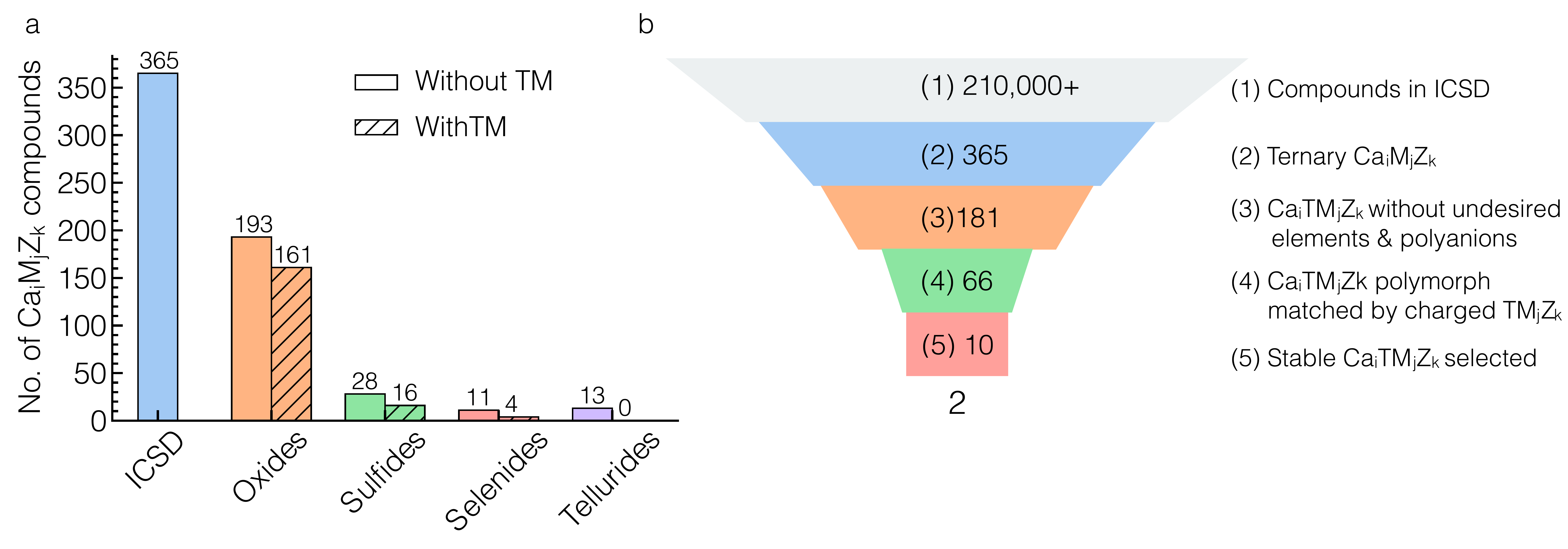}
\caption{
(a) Distribution of ternary {\ch{Ca_iM_jZ_k}} compounds, where {\ch{Z}} = O, S, Se, or Te and \ch{M} is a non-Ca cation. Hatched bars indicate  {\ch{Ca_iTM_jZ_k}}-like compounds with a transition metal ({\ch{TM}}) of the $d$-block and $Z$ any chalcogen element, i.e., O, S, Se or Te. (b) Filtering procedure to identify ternary Ca-containing electrodes with composition of \ch{Ca_iTM_jZ_k}, where {\ch{TM}} is redox-active. The number in each coloured panel identifies the number of accepted materials at each stage of the screening.
}
\label{fig:sifting} 
\end{figure*}

\noindent Figures~\ref{fig:sifting}a shows the distribution of Ca-containing ternary compounds, with prototype formula \ch{Ca_iM_jZ_k}, where \ch{M} is a generic cation (and not necessarily an open-shell transition metal) and \ch{Z} is the anion species, as obtained from the inorganic crystal structure database (ICSD).\cite{Bergerhoff1983,Hellenbrandt2004}  We excluded structures with non-Ca cations that can potentially exhibit electrochemical activity, such as Li, Na, Mg, Zn, K, Rb, and Cs, from the list of cations considered as \ch{M}. In the reminder of this study we will address as chalcogenides all compounds with formula {\ch{Ca_iM_jZ_k}} where  {\ch{Z}} can be O, S, Se, or Te. Of the more than 210,000 structures in ICSD, only 412 satisfy the compositional constraint of {\ch{Ca_iM_jZ_k}}, with Z = O, S, Se, or Te. Furthermore, a significant portion of identified \ch{Ca_iM_jZ_k} are unsurprisingly oxides\cite{Brown1988,Yang2014} (orange bars in Figure~\ref{fig:sifting}a), with a considerably lower number of other chalcogenides. 

The hatched bars  in Figure~\ref{fig:sifting}a represent {\ch{Ca_iTM_jZ_k}} compounds with a transition metal {\ch{TM}} of the $d$-block and $Z$ any chalcogen element, i.e., O, S, Se or Te. Therefore, the hatched bars exclude mixed anion moieties, such as {\ch{SiO_4^{4-}}}, {\ch{PO_4^{3-}}}, etc. Figure~\ref{fig:sifting}b presents our search strategy to identify potential Ca conductors/electrodes, with chemical formula \ch{Ca_iTM_jZ_k}. We utilised the combined information available in the ICSD (i.e., structures and characterization conditions),{\cite{Hellenbrandt2004}} and the Materials Project (MP; thermodynamic stabilities)\cite{Jain2013} to arrive at our candidates. The screening procedure, where we induce constraints on composition, redox-activity, charge-balance, and thermodynamic stability, followed the steps below: 
\begin{enumerate}

\item \emph{Identification of {\ch{Ca_iM_jZ_k}} compounds in the ICSD.} This initial screen could  identify 412 oxides or chalcogenides {\ch{Ca_iM_jZ_k}} materials as shown in Figure~{\ref{fig:sifting}}a.

\item  \emph{Selection of {\ch{Ca_iTM_jZ_k}} compounds with a transition metal  {\ch{TM}} of the d-block}, where the TM can be redox-active. For example, even though Sc and Zn are \emph{d}-block TMs, they are not known to be redox-active.
 As shown in Figure~{\ref{fig:sifting}}b, this important filter could further reduce our dataset to 181 {\ch{Ca_iTM_jZ_k}} candidates, of which only 161 are oxides, 16 are sulfides and 4 are selenides (see Figure~{\ref{fig:sifting}}b). Furthermore, all 13 telluride {\ch{Ca_iM_jZ_k}} compounds were entirely made by mixed anion species (e.g., {\ch{TeO_6^{2-}}}) and are eliminated from our selection. In this step we also eliminated all {\ch{Ca_iTM_jZ_k}} that were characterized under non-ambient conditions.

\item \emph{Identification of discharged compositions {\ch{Ca_iTM_jZ_k}}, which can be matched by charge-balanced decalciated compositions {\ch{TM_jZ_k}}}. As Ca is deintercalated from a \ch{Ca_iTM_jZ_k}, either partially or fully, it is important that the \ch{TM} species exhibits oxidation states that are physically plausible, hence ensuring charge-neutrality of the deintercalated composition. For example, {\ch{CaVO3}} is a well-known Ca-containing orthorhombic perovskite that satisfies the chemical formula of {\ch{Ca_iTM_jZ_k}} and contains an open-shell transition metal (i.e., {\ch{V^{4+}}}). However, the fully-charged version of {\ch{CaVO3}}, namely, {\ch{VO3}}, requires V to exhibit an oxidation state of 6+, which is not physical. Thus, {\ch{CaVO3}} does not qualify as a candidate across this filter. Note that charge-neutrality is one of the necessary, but not sufficient, conditions for thermodynamic stability. Hence, during this filtering process, we tried to identify possible charge-neutral deintercalated composition(s) for every \ch{Ca_iTM_jZ_k} considered. In particular, we checked the availability of ordered structures of both fully ({\ch{TM_jZ_k}}) and partially ({\ch{Ca_{l<i}TM_jZ_k}}) deintercalated compositions that were charge-neutral in the ICSD as well as Materials Project. Although any of the screened Ca chemistries may allow partial decalciation, we preferred fully charged structures in all possible cases, since that maximizes theoretical capacities. We only considered partially decalciated compositions when ordered structure(s) were available at such compositions: else, the computational costs become intractable. Specifically, only two specific cases of partial Ca extraction came through our search, namely, {\ch{CaV2O6}} and {\ch{Ca3Cu6O12}}. Also, for the two eventual candidates proposed in this work, {\ch{CaNb2O4}} and {\ch{CaV2O4}}, we did consider some Ca configurations at partial decalciation, as illustrated by the voltage curves plotted in Figure~S1 of the Supporting Information (SI). Wherever possible, we have identified discharge/charge structure pairs that would enable a topotactic Ca (de)intercalation. If the charged composition exhibits multiple polymorphs, we only consider the structure that topotactically matches the discharged structure and the ground state structure (if different from topotactically-matched structure) for further calculations. After this step, the number of possible {\ch{Ca_iTM_jZ_k}} electrodes further reduced to 66.

\item \emph{Identification of  final set of 10 materials to investigate with first-principles calculations.} We arrived at our final set by eliminating all the \ch{Ca_iTM_jZ_k} structures which appeared thermodynamically unstable, as quantified by the 0 K energy above the convex hull (E$^\mathrm{hull}$) that is available on MP, on their respective compositional phased diagrams. We assumed the \ch{Ca_iTM_jZ_k} compounds to be unstable if their E$^\mathrm{hull}>$~30~meV atom$^{-1}$, which has been widely accepted as a rule-of-thumb in stability-based screening studies.\cite{Sun2016} Importantly, the availability of a thermodynamically stable \ch{Ca_iTM_jZ_k} compound ensures that there is no thermodynamic driving force for forming detrimental conversion products, instead of Ca-intercalation, upon reduction of the charged \ch{TM_jZ_k} structure with Ca (see Table~\ref{tbl:itercalationdata} and Sec.~\ref{sec:methods} for further discussion).\cite{Canepa2017,Hannah2018} Also, we did not impose any thermodynamic stability restriction on the deintercalated {\ch{TM_jZ_k}} compositions (see discussion in Sec.~{\ref{sec:ca-intercalation}}). 
 
\end{enumerate}
Of the 412 \ch{Ca_iM_jZ_k}  oxides or chalcogenides, we fully investigated the thermodynamic and kinetic properties of ten candidates and subsequently identified two structures as promising Ca conductors/electrode materials that merit further experimental investigation, namely \ch{CaV2O4} and \ch{CaNb2O4}. A comprehensive list of all the materials investigated in this study is provided in Table~S1 of the SI.

\subsection{Ca Intercalation Characteristics in \ch{Ca_iTM_jZ_k}  Materials}
\label{sec:ca-intercalation}

Table~\ref{tbl:itercalationdata} summarizes the intercalation characteristics of the 10 \ch{Ca_iTM_jZ_k} candidates, including the space groups, the ICSD or MP IDs, and E$^\mathrm{hull}$ of the discharged and charged states, the average Ca topotactic intercalation voltage (Eq.~\ref{eq:voltage}), the gravimetric capacity, and the specific energy (calculated from the theoretical average voltage) for all charged-discharged pairs.  Table~\ref{tbl:itercalationdata} also contains the Ca coordination number and the Ca $E_m$ in the discharged state, as well as the current cost of the raw TMs in the \ch{Ca_iTM_jZ_k} compounds. We included \ch{CaV2O5} as the 11$^\mathrm{th}$ compound in the list as a benchmark for our computational method. Overall, the screened set of ten compounds in Table~{\ref{tbl:itercalationdata}} consists of only two sulfides (\ch{CaCu2S2}-\ch{Cu2S2} and \ch{CaMo6S8}-\ch{Mo6S8}), with the rest being oxides, testament to the significantly larger collection of available oxides that naturally form with Ca in preference to other chalcogenides. All $E_m$ reported in Table~{\ref{tbl:itercalationdata}} are calculated to represent the discharged structure, i.e., in the dilute Ca-vacancy limit. Also, Table~{\ref{tbl:itercalationdata}} includes both topotactic (T) and non-topotactic (N) average intercalation voltages, where relevant.

\begin{table*}[ht!]
  \caption{
  Computed characteristics of the 10 discharged \ch{Ca_iTM_jZ_k} and charged \ch{TM_jZ_k} compounds resulting from our screening.  The ICSD or MP ID of each material is provided. E$^\mathrm{hull}$, Voltage, Capacity (Cap.), Specific energy (Spec.) and cost of transition metal are reported in units of meV/atom, Volts vs.\ \ch{Ca}/\ch{Ca^{2+}}, mAh/g, Wh/kg and \$/kg, respectively.  Mech. indicate whether Ca intercalation follows a topotactic (T) or a non-topotactic (N) mechanism. The average coordination number (CN) of \ch{Ca^{2+}} and the computed $E_m$ (in meV) in the discharged structures (i.e., dilute vacancy limit) are also reported.  }
  {\scriptsize
  \label{tbl:itercalationdata}
  \begin{tabular*}{\textwidth}{@{\extracolsep{\fill}}lllclclllcl@{}}

    \hline \hline \\[-7pt]
    {\bf Material}     & {\bf ICSD/MP} & {\bf Space Group}     &  {\bf Mech.}    &  {\bf E$\mathbf {^{hull}}$} &  {\bf CN}   & $\mathbf{E_m}$  & {\bf Voltage}              & {\bf Cap.}             & {\bf Spec.}       & {\bf Cost} \\[2pt]
    \hline \\[-7pt]
    \ch{CaV3O7}  & 2507  & $Pnma$      &--    & 2            &   7   & 2,892  & -- & \multirow{3}{*}{176} & -- &  \multirow{3}{*}{17.40$^\mathrm{a}$}\\ 
    \ch{V3O7}    & mvc-13330& $Pnma$   & T     & 51       &  --  & --    &    3.34  &  &587 \\
    \ch{V3O7}  & mp-622640 & $C2/c$   & N & 8 & -- &  --  & 3.11 & & 547\\
    \hline \\[-7pt]
    \ch{Ca2V2O6} & 237336  &  $Pnma$  & --& 0           &   7   & 3,235& -- & \multirow{3}{*}{193} & -- & \multirow{3}{*}{17.40$^\mathrm{a}$} \\ 
    \ch{CaV2O6}  & This work$^\mathrm{d}$  & $Pnma \rightarrow Pmn2_1$        & T &      95       & --       &  --   & 2.96 & & 571\\
    \ch{CaV2O6}  & 21064  & $C2/m$        & N & 0            & --       &  --   &   2.46 & & 474 \\
    \hline \\[-7pt]
    \ch{CaV2O4} &  164185 & $Pnma$         & --&  0             & 8    &654  & -- & \multirow{3}{*}{260} &   -- & \multirow{3}{*}{17.40$^\mathrm{a}$}\\ 
    \ch{V2O4}   &  mp-777479 & $Pnma$       &  T &26           & --       &  --   & 2.40 & & 624 \\
    \ch{V2O4}   & 10141 & $P4_2/mnm$  &  N &0 & -- &  --   & 2.34 & &608 \\

     \hline  \\[-7pt]
    \ch{CaNb2O4}  & 88779  & $Pbcm$        & --& 0        & 6        & 785     & -- & \multirow{3}{*}{185} & -- & \multirow{3}{*}{30.20$^\mathrm{b}$}\\ 
    \ch{Nb2O4}    & This work$^\mathrm{d}$  & $Pbcm \rightarrow Pmma$   &  T     & 308        & --          &  --   & 2.71 & & 501 \\
    \ch{Nb2O4}    & 96     & $I4_1/a$      &  N &0         & --          &  --   & 1.78 & & 329 \\
    
    \hline \\[-7pt]
    \ch{Ca5Cu6O12} & 91059& $P2_1/c$   &  --       & 28         & 6      & --   &-- & \multirow{3}{*}{139} & -- &  \multirow{3}{*}{ 8.75$^\mathrm{c}$} \\ 
    \ch{Ca3Cu6O12} & This work$^\mathrm{d}$ & $P2_1/c \rightarrow P2_1$  &   T &  92          & -- &  --   & 3.19 & & 443\\
    \ch{Ca3Cu6O12} & mp-1540145& $I4_1/a$   &  N & 0           & --        &  --   & 2.83 & &393 \\
    \hline \\[-7pt]
    \ch{CaMoO3}  &  246082& $P2_1/c$       &   -- &  0           & 8        &  2,072 & -- & \multirow{3}{*}{291} & --  &   \multirow{3}{*}{24.75$^\mathrm{c}$} \\ 
    \ch{MoO3}    &  mp-18856  & $P2_1/c$      &  T & 0           & --        &  --   & 2.77 &  & 806 \\
    \ch{MoO3}    & 35076 & $Pnma$                 &  N & 39        &  --       &  --   & 2.76 & & 803\\  
    \hline \\[-7pt]
    \ch{CaIrO3}  &  420479 & $Cmcm$         & \multirow{2}{*}{T} &  0           & 8     & \multirow{2}{*}{1,219}    & \multirow{2}{*}{3.29} & \multirow{2}{*}{191} & \multirow{2}{*}{629} &   \multirow{2}{*}{193,000.00$^\mathrm{c}$}  \\ 
    \ch{IrO3}    &  mp-1097041 & $Cmcm$    &  & 0           & --        &        & \\
    \hline \\[-7pt]
    \ch{CaRh2O4}  & 170597 & $Pnma$        & -- & 0         &8        & 1,110  &  -- & \multirow{3}{*}{173} & -- &  \multirow{3}{*}{860,000.00$^\mathrm{c}$}    \\ 
     \ch{Rh2O4}    & This work  &        $Pnma$   & T            & 67        & --         &   --   &2.84  & & 491 \\
    \ch{Rh2O4}    & 28498  & $P4_2/mnm$    & N & 0          & --         &  --   & 2.63& & 455 \\
    \hline \\[-7pt]
    \ch{CaCu2S2} & 241336 & $P\bar{3}m1$   & -- & 0          & 6       & 1,622   & --  & \multirow{3}{*}{232} & -- &  \multirow{3}{*}{8.75$^\mathrm{c}$}\\ 
        \ch{Cu2S2}   & This work$^\mathrm{d}$   & $P\bar{3}m1 \rightarrow C2/m$     & T &  108        & --          &  --   & 2.17 & & 503 \\
    \ch{Cu2S2}   & 63328  & $Cmcm$         & N & 0         & --          &   --   &1.95 & & 452 \\

    \hline \\[-7pt]
    \ch{CaMo6S8} & 619423 & $R\bar{3}$    & \multirow{2}{*}{T} & 17       & 8       & --     &  \multirow{2}{*}{1.82} & \multirow{2}{*}{61} & \multirow{2}{*}{112} & \multirow{2}{*}{24.75$^\mathrm{c}$} \\ 
    \ch{Mo6S8}   & 86788  & $R\bar{3}$    &  & 66         & --         &  --   \\
        \hline \\[-7pt]
    $\alpha$-\ch{CaV2O5}$^e$ &  82689  & $Pmmn$     &\multirow{2}{*}{T} &     0          & 8   & \multirow{2}{*}{1,869}            & \multirow{2}{*}{3.13}& \multirow{2}{*}{241} & \multirow{2}{*}{757} &\multirow{2}{*}{17.40$^\mathrm{a}$} \\ 
    \ch{V2O5}   &  15798  & $Pmmn$          &  & 0    &--       \\[2pt]
    \hline \hline
  \end{tabular*}
    \begin{flushleft}
   $^\mathrm{a}$Vanadium price obtained from \url{https://www.vanadiumprice.com}  and sold as  \ch{V_2O_5}.\\
   $^\mathrm{b}$Niobium price obtained from \url{https://www.niobiumprice.com} and sold as  \ch{Nb_2O_5}. \\
   $^\mathrm{c}$Metal prices obtained from \url{https://www.dailymetalprice.com/}. \\
   $^\mathrm{d}$Upon topotactic extraction of {\ch{Ca^{2+}}}, the symmetry of the charged structure becomes different from the discharged structure. \\
   $^\mathrm{e}$Considered as a calibration compound.
   \end{flushleft}
  }
  \end{table*}

A quick glance of Table~\ref{tbl:itercalationdata} reveals that all but four discharged compounds exhibit E$^\mathrm{hull}$ =~0~meV/atom. Three of the metastable \ch{Ca_iTM_jZ_k} compounds show a E$^\mathrm{hull}<~$30~meV/atom (i.e., \ch{CaV3O7}, \ch{Ca5Cu6O12} and \ch{CaMo6S8}).  Among the 10 discharged-charged pairs considered, the charged compounds can be metastable in their respective phase diagrams.  For example, the deintercalated Chevrel-phase is theoretically expected to decompose as {\ch{Mo6S8}} $\rightarrow$ {2\ch{MoS2}}~+~{\ch{Mo}}  ($\sim$--88.0~kJ~mol$^{-1}$ and --908~meV). However, the Chevrel cathode has shown reversible intercalation of various ions over several cycles,{\cite{Levi2002,Levi2005,Levi2010}} indicating that the charged Chevrel phase is kinetically stabilized in such cases. Nevertheless, reversible Ca intercalation into the Chevrel cathode has not been yet reported so far experimentally despite previous theoretical works{\cite{Smeu2016,Smeu2017}} comparing the intercalation behavior of Ca to other ions, suggesting possible instability associated with topotactic Ca insertion into the Chevrel. In any case, the thermodynamic metastability of charged compounds may not represent a significant deterrent for the reversible operation of a Ca battery or the ability to experimentally synthesize such compounds.

Unsurprisingly, from Table~\ref{tbl:itercalationdata}, we realize that the largest intercalation voltages are achieved by Vanadium-containing phases, which can be attributed to higher (more positive) standard reduction potential for V (e.g., \ch{V^{IV}/V^{III}} +0.34 V vs.\ SHE) compared to the other transition metals considered (e.g., \ch{Mo^{IV}/Mo^{III}} --0.01 V vs.\ SHE).\cite{CRCHandbook2007} Indeed, Ca intercalates in \ch{V3O7} and $\alpha$-\ch{CaV2O5} with $\sim$3.34 and $\sim$3.13~V vs.\ \ch{Ca/Ca^{2+}}, respectively, significantly higher than \ch{MoO3} ($\sim$2.77~V), \ch{Nb2O4} ($\sim$1.79~V), \ch{Cu6O12} ($\sim$2.83~V), or \ch{Rh2O4} ($\sim$2.63~V). The lower voltage exhibited by \ch{V2O4} ($\sim$2.40~V) relative to the \ch{V3O7} and \ch{V2O5} frameworks can be attributed to the lower (more negative) reduction potential of the  \ch{V^{IV}/V^{III}} than \ch{V^{V}/V^{IV}}.  

The predicted Ca intercalation voltage in \ch{V2O5} is in excellent agreement with previous reports,\cite{Carrasco2014,Gautam2015,Das2021}  providing credibility to our methodology. The  large Ca intercalation voltages and the accessibility to several stable oxidation states makes the Vanadium phases the most attractive in terms of gravimetric capacities and specific energy density, as shown in Table~\ref{tbl:itercalationdata}.  Also, our screening reveals other promising, previously unreported phases, including \emph{i}) the CaIrO$_3$ ($\sim$3.29~V vs.\ \ch{Ca/Ca^{2+}}), \emph{ii}) \ch{Ca5Cu6O12}, and \emph{iii}) \ch{CaRh2O4}. We do not expect Ir and Rh compounds to be competitive with V compounds due to the significantly high cost of the raw metals.

\begin{figure}[ht!] 
\includegraphics[width=\columnwidth]{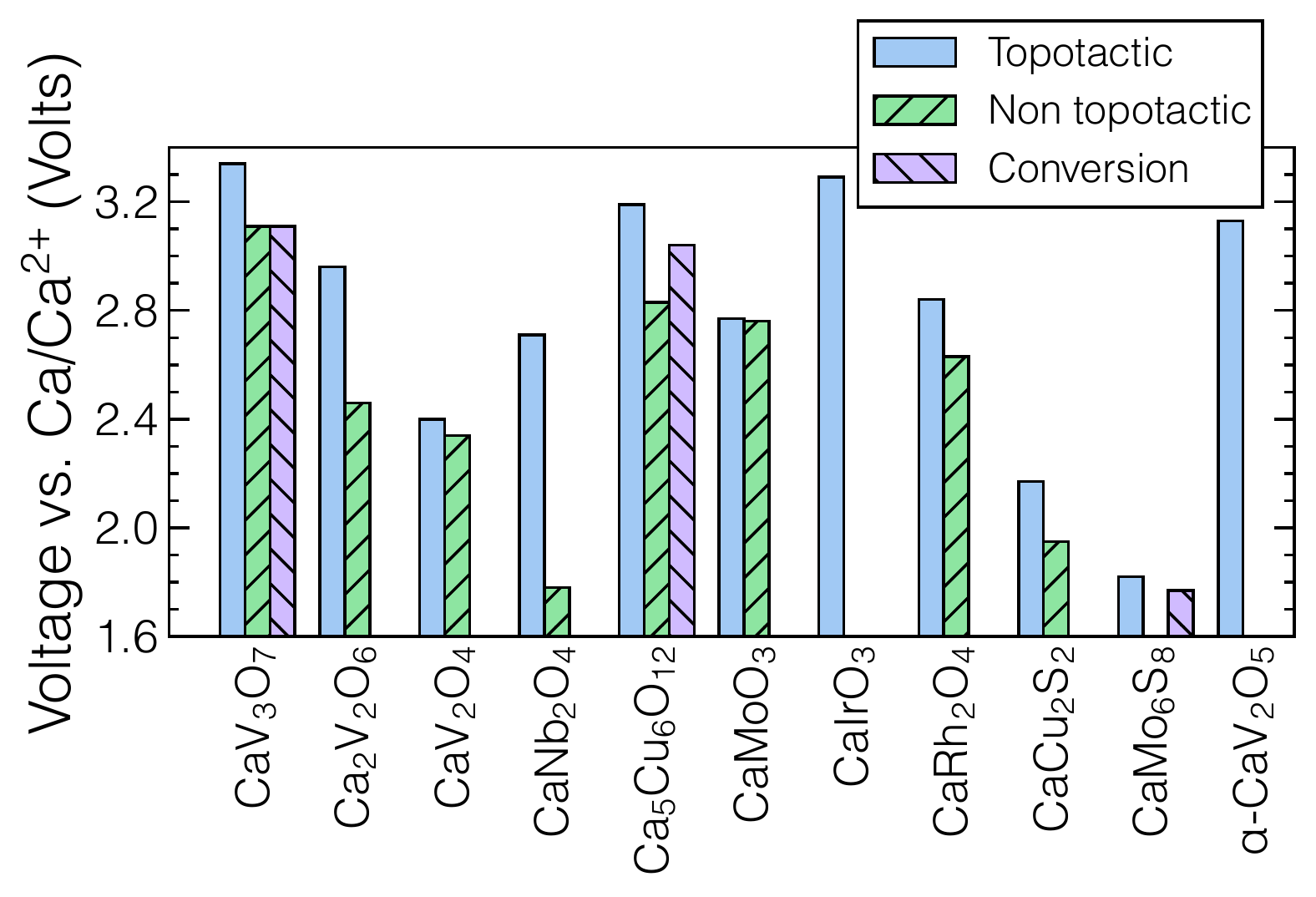}
\caption{
Computed intercalation voltages vs.\ \ch{Ca}/\ch{Ca^{2+}} of compounds in Table~{\ref{tbl:itercalationdata}}. Light-blue and hatched green bars indicate voltages associated with topotactic and non-topotactic intercalation, respectively. Missing bars indicate the unavailability of  non-topotactic \ch{TM_jZ_k} compounds in the ICSD or MP.  The competitive conversion reactions are shown as purple hatched bars.  The conversion voltage of the Chevrel phase  {\ch{Mo6S8}}$+${2\ch{Ca}}~$\rightarrow$~{3\ch{MoS2}}$+${3\ch{Mo}}$+${3\ch{CaS}}  is comparable to the topotactic insertion voltage, which is due to {\ch{Mo6S8}} being thermodynamically unstable ($\sim$66~meV E$^\mathrm{hull}$, Table~{\ref{tbl:itercalationdata}}).
}  
\label{fig:voltages} 
\end{figure}

During (de)intercalation, if the host framework (i.e., \ch{TM_jZ_k}) does not phase transform, resulting in the symmetry of the host lattice being preserved, the (de)intercalation process is said to be topotactic. However, several cathode materials, among those that we have considered, are expected to undergo phase transformations upon Ca deintercalation, which can result in a different structure (and/or symmetry) of the charged \ch{TM_jZ_k} compared to the discharged \ch{Ca_iTM_jZ_k} compound.  Upon Ca extraction \ch{CaNb2O4} is thermodynamically driven to transform from the orthorhombic (space group:~$Pbcm$) to  tetragonal ($I4_1/a$) (see Table~\ref{tbl:itercalationdata}). 
For battery operation, topotactic intercalation mechanisms are preferred since phase transformations can lead to undesired straining (and/or fracture) of cathode particles, resulting in impedance build-up and even loss of electrochemical activity. In general, non-topotactic intercalation mechanisms are caused by the high instability of the topotactic charged phase (e.g., $\delta$-\ch{V2O5} for Ca) with respect to the corresponding (more stable) non-topotactic charged phase ($\alpha$-\ch{V2O5} for Ca). Hence, the lower the difference in stability between the topotactic and non-topotactic phases, which can be quantified by the difference between the topotactic and non-topotactic voltages, the lower the tendency for a host framework to adopt a non-topotactic mechanism.

We have highlighted the difference between the topotactic (light-blue bar) and non-topotactic (green hatched bar) computed intercalation voltages for all ten identified candidates (and $\alpha$-\ch{V2O5}) in Figure~\ref{fig:voltages}. Importantly, the largest difference between topotactic and non-topotactic voltage is observed in {\ch{CaNb2O4}}$\rightarrow${\ch{Nb2O4}} of $\sim$1.0~V, with other compounds showing smaller differences between the two voltages. Note that even in frameworks presenting a significant difference between the topotactic and non-topotactic voltages (i.e., a few tenths of a V), the topotactic mechanism may remain active since, this process is typically  kinetically favoured over  the  non-topotactic intercalation. For example, the $Pnma$ phase of {\ch{CaV2O4}} is experimentally reported, but the charged $Pnma$-{\ch{V2O4}} phase is metastable (by $\sim$26~meV/atom) than the stable $P4_2/mnm$-\ch{V2O4} phase, leading to a topotactic (non-topotactic) voltage of $\sim$2.40~V ($\sim$2.34~V) vs.\ {\ch{Ca}}/{\ch{Ca^{2+}}} (see Figure~{\ref{fig:voltages}}). Given the small change between the topotactic and non-toptactic voltages for \ch{V2O4}, we speculate that the $Pnma$-{\ch{V2O4}} polymorph may benefit from kinetic stabilization. The conversion voltage upon discharge in \ch{Ca3Cu6O12} (purple bar), is higher than the non-topotactic intercalation voltage, in agreement with the metastability of the discharged \ch{Ca5Cu6O12} structure (Table~\ref{tbl:itercalationdata}). In the case of {\ch{CaV3O7}}, the conversion voltage is only marginally higher than non-topotactic intercalation voltage, as indicated by the marginal instability of {\ch{V3O7}} (E$^\mathrm{hull}\sim$2~meV/atom). However, given the larger topotactic intercalation voltage, we can expect the topotactic intercalation reaction to be kinetically favored in both the \ch{Ca5Cu6O12}-\ch{Ca3Cu6O12} and {\ch{CaV3O7}}-{\ch{V3O7}} systems. Nevertheless, topotactic (de)intercalation in multivalent systems are in general more complex than monovalent (Li/Na) systems,{\cite{Dompablo2019,Canepa2017}} as highlighted by recent studies that have demonstrated decomposition reactions upon Ca extraction in {\ch{CaTaN2}},{\cite{Verrelli2019}} and phase transformations upon Ca removal in {\ch{Ca3Co2O6}}.{\cite{Tchitchekova2018a}}

\subsection{\ch{Ca^{2+}} Migration Barriers}

Excluding \ch{Mo6S8} that has  already been studied extensively as a multivalent cathode,\cite{Canepa2017} we computed the migration barriers for Ca$^{2+}$ $E_m$ for the remaining 9 materials of Table~\ref{tbl:itercalationdata}.  All computed migration energy paths shown in Section~S4 of the SI. We have calculated the $E_m$ for all materials assuming the dilute vacancy limit, where Ca migration occurs via a vacancy-migration mechanism, unless otherwise specified. Despite our best efforts, we were unable to converge the $E_m$ for \ch{Ca5Cu6O12} structure and thus have not reported its $E_m$ in this work. Notably, our computed $E_m$ of 1869~meV in $\alpha$-\ch{CaV2O5} (see Figure~S10 in  SI) is in excellent agreement with a previous calculation ($\sim$1900~meV) done in the dilute vacancy limit.\cite{Gautam2015} 

\begin{figure}[h!] 
\includegraphics[width=\columnwidth]{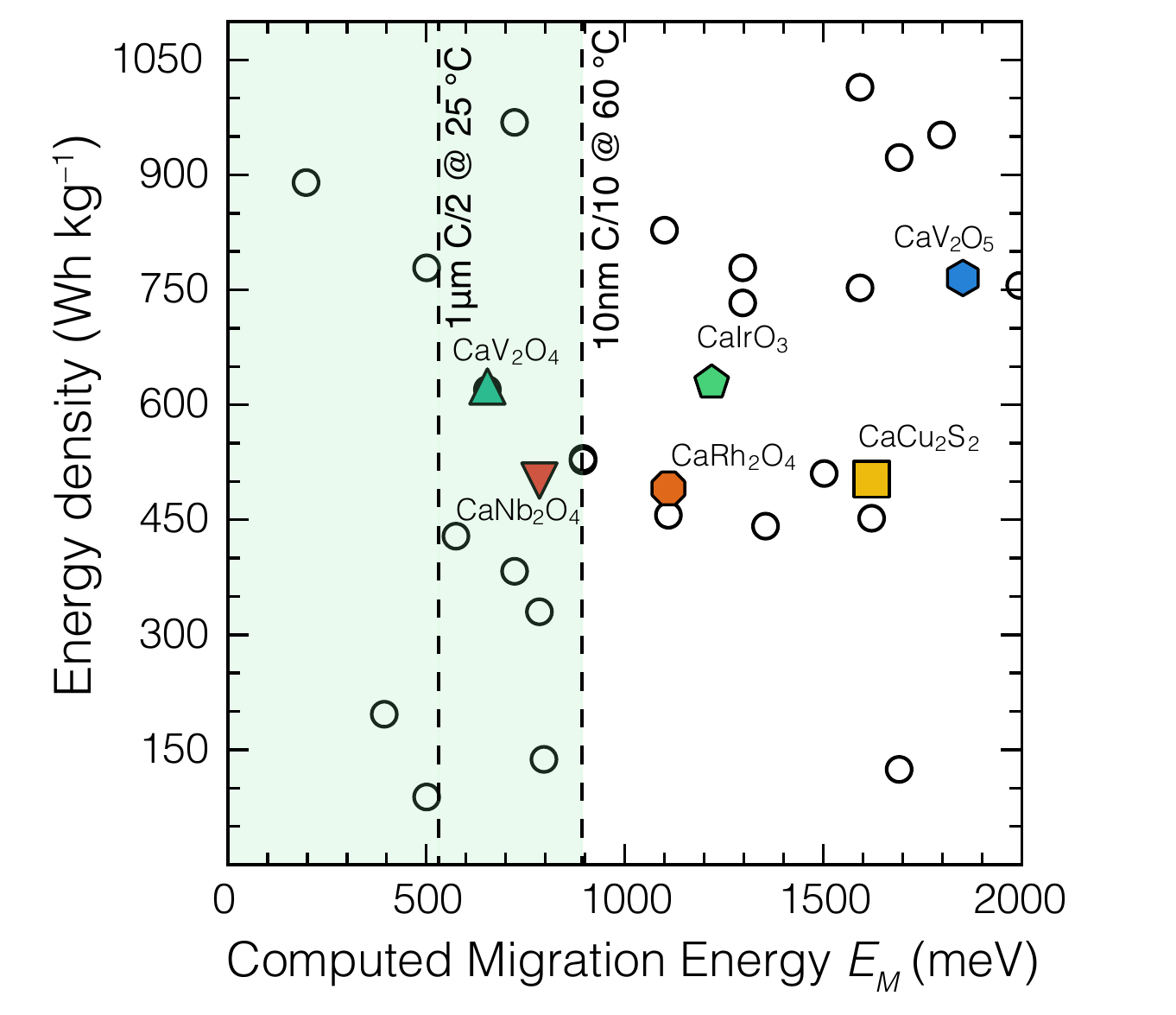}
\caption{
Computed {\ch{Ca^{2+}}} migration barriers (in meV) plotted with respect to the predicted specific energies (in Wh~kg$^{-1}$). Coloured points 
are  materials investigated in this study, empty  circles are computationally predicted data points from  Ref.~\citenum{Dompablo2019}.  The green ribbon shows favourable {\ch{Ca^{2+}}} migration barriers. Reported energy densities rely on topotactic intercalation voltages, which indicate the most probable Ca insertion reactions. Specific tolerable migration barriers  (i.e., 525 and 895~meV) at specific (dis)charge conditions are marked by dashed lines.\cite{Rong2015,Canepa2017,Chen2019}  Materials considered in our study that exhibit barriers above 2000~meV (\ch{CaV3O7}, \ch{Ca2V2O6}, and \ch{CaMoO3}) are not shown in the figure.}  
\label{fig:barriers} 
\end{figure}

Figure~\ref{fig:barriers} shows the computed Ca migration barriers in selected materials (see coloured shapes) vs.\ their predicted specific energies (y-axis). Black empty circles in Figure~\ref{fig:barriers} indicate previously reported $E_m$ for different materials from Arroyo-de Dompablo and collaborators,\cite{Dompablo2019} which serve as useful reference points for our candidates. The green region in Figure~{\ref{fig:barriers}} marks the range of  tolerable migration barriers $E_m$. In Figure~{\ref{fig:barriers}}, dashed lines at $\sim$525~meV and at $\sim$895~meV correspond to Ca (de)intercalation across a micron-sized cathode particle at 298~K and C/2 rate, and Ca (de)intercalation across a 10~nm cathode particle, at 333~K ($\sim$60 ${^{\circ}}$C) and C/6 rate (6 hours), respectively.{\cite{Rong2015,Canepa2017,Chen2019}} The variation of the maximum tolerable $E_m$ vs.\ particle size, C-rate, and temperature is given in Figure~S2 of the SI.

Three main observations can be made from Figure~{\ref{fig:barriers}} and Table~{\ref{tbl:itercalationdata}}: \emph{i}) the computed  barriers for all these materials are found to be larger than the $\sim$650~meV threshold of Rong et al.\cite{Rong2015}, with \ch{Ca2V2O6} displaying the largest $E_m$ of $\sim$3235~meV,  \emph{ii}) of the ten materials investigated, only two materials \ch{CaNb2O4} ($\sim$785~meV) and \ch{CaV2O4} ($\sim$654~meV) qualify the mobility criterion since their predicted $E_m$ are under the highest threshold of 895~meV, and \emph{iii}) the specific energy of \ch{CaV2O4} is quite competitive with other compounds previously reported by Arroyo-de Dompablo et al.,\cite{Dompablo2019} and far exceeds commercial Li-ion cathodes.\cite{Whittingham2014}  Although eight out of the ten shortlisted candidates display exceedingly large \ch{Ca^{2+}} migration barriers, we believe there is value in communicating these high $E_m$ values to not only focus experimental effort on the promising candidates but also to reformulate the design rules that can aid in designing better Ca-conductors. 

\subsection{Assessment of \ch{CaV2O4} and \ch{CaNb2O4} electrodes}


\begin{figure*}[ht!] 
\includegraphics[width=\textwidth]{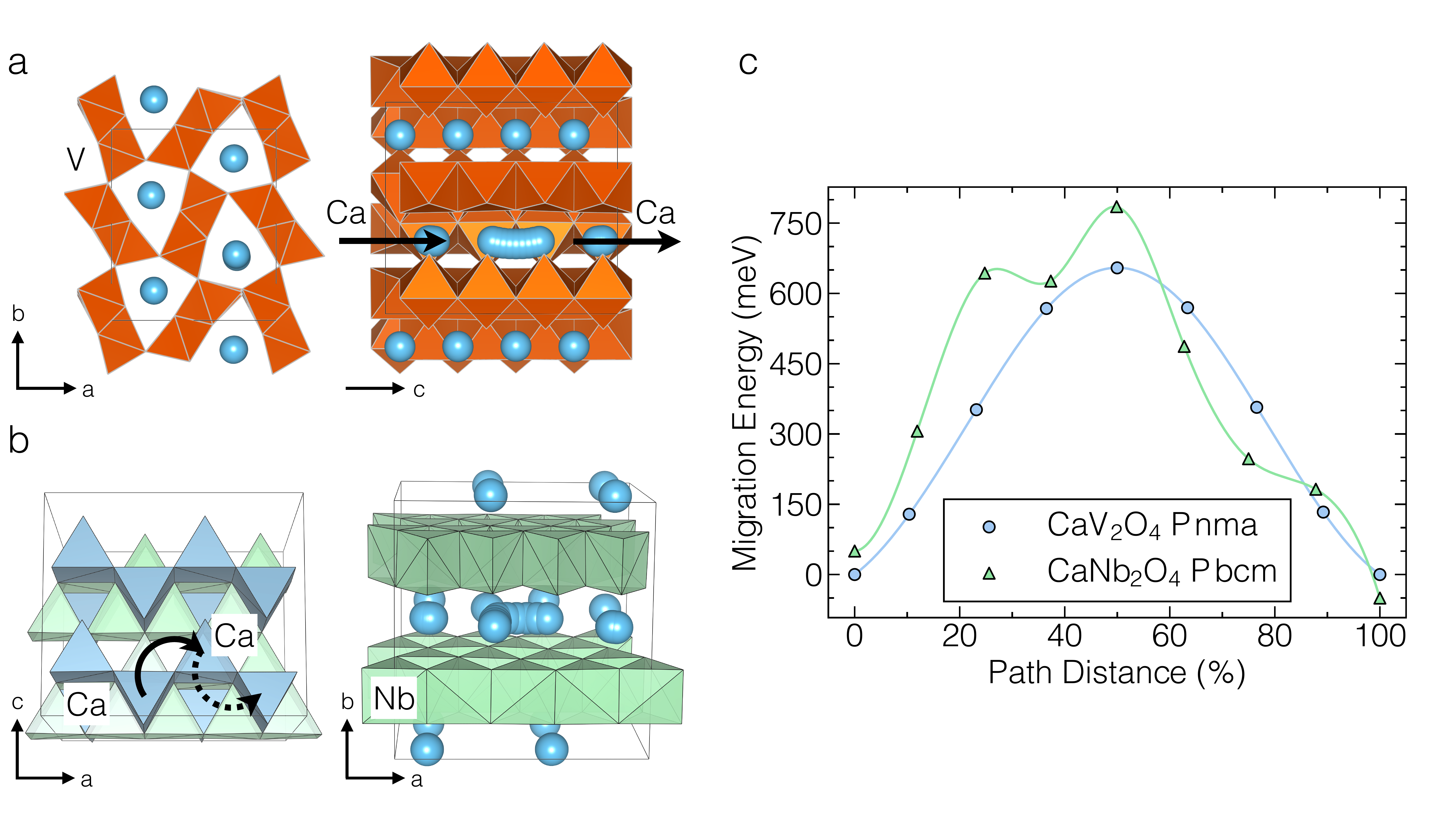}
\caption{
Panels (a) and (b) show the structural features of \ch{CaV2O4} (with space group:~$Pnma$) and \ch{CaNb2O4} ($Pbcm$) highlighting the \ch{Ca^{2+}} migration path. V and Nb polyhedra are depicted in orange and green, respectively, while Ca is shown either as blue spheres or polyhedra. Arrows indicates the direction of \ch{Ca^{2+}} migration. (c) Computed migration energy paths (in meV) of \ch{CaV2O4}  (blue circles) and \ch{CaNb2O4} (green triangles) vs.\ \ch{Ca^{2+}} displacement. 
}  
\label{fig:specbarrier} 
\end{figure*} 

\noindent {\bf \emph{Pnma}-\ch{CaV2O4}, post-spinel:} The structure of \ch{CaV2O4} arranges in the post-spinel-type prototype  (also referred as the calcium ferrite structure), where the vanadium resides in distorted  edge-sharing and corner-sharing \ch{VO6} octahedral units (Figure~{\ref{fig:specbarrier}}a).  The post-spinel structure has been described in detail elsewhere,\cite{Hannah2017} where Mg occupies an 8-coordinated site in  the post-spinel \ch{Mg_{x}V_{2-y}Ti_{y}O4}. Similarly, \ch{CaV2O4} exhibits \ch{Ca^{2+}} in an 8-coordinated environment with Ca-migration occurring  along 1D channels in the $c$ direction (see Figure~\ref{fig:specbarrier}a). Therefore, $Pnma$-\ch{CaV2O4} is a 1D conductor, leading to a migration topology of 8$\rightarrow$5$\rightarrow$3$\rightarrow$5$\rightarrow$8 between two equivalent Ca sites. The migration energy path of Figure~\ref{fig:specbarrier}c reaches its maximum ($\sim$654~meV) when Ca is in a 3-coordinated (triangular) site. Post-spinel structures have been previously considered for Ca intercalation. For example, {\citeauthor{Dompablo2016}}{\cite{Dompablo2016}} have studied Ca intercalation in the {\ch{CaMn2O4}} (CM)- and {\ch{CaFe2O4}} (CF)-type post-spinel structures of {\ch{CaMn2O4}} composition.
 \\  

\noindent {\bf \emph{Pbcm}-\ch{CaNb2O4}, layered:} In \ch{CaNb2O4}, Nb atoms are  6-coordinated by oxygens forming edge-sharing, distorted  triangular \ch{NbO6} prisms, which give rise to the layered structure in Figure~\ref{fig:specbarrier}b. \ch{Ca^{2+}}-ions occupy an ordered configuration of 6-coordinated irregular prismatic sites residing in-between the \ch{NbO6} layers. Therefore, \ch{CaNb2O4} can be considered as a 2D-conductor. The in-layer migration of \ch{Ca^{2+}} follows a topology 6$\rightarrow$4$\rightarrow$6, with a maxima of $\sim$785~meV at the 4-coordinated (rectangular) site, as indicated by Figure~\ref{fig:specbarrier}c.  As  highlighted  by the solid and dash curved arrows of Figure~{\ref{fig:specbarrier}}b,  we have calculated the migration barrier from half the total path, which explains the difference between the energies of the ground-state structures (at 0 and 100~\%, Figure~{\ref{fig:specbarrier}}c).


\section{Discussion}
\label{sec:discussion}

So far, the identification of promising intercalation cathode  materials has been limited by the relatively large ionic size of \ch{Ca^{2+}}, which decreases the availability of structural framework compatible with hosting and transporting \ch{Ca^{2+}} ions.\cite{Koettgen2020}  In this report, we have searched the chemical space of  ternary {\ch{Ca_iTM_jZ_k}}  compositions using a combination of existing databases, intuitive filtering criteria, and DFT calculations. From an initial pool of 412 Ca-containing materials found in the ICSD, we shortlisted 10 promising candidates, whose intercalation properties were assessed further via \emph{ab initio} methods.  Our screening strategy (of Figure~\ref{tbl:itercalationdata}) is robust as it also provides known candidates that have been previously studied experimentally (and computationally) as Ca-cathodes. While {\ch{TiS_2}} has been demonstrated to intercalate reversibly  {\ch{Ca^{2+}}},{\cite{Tchitchekova2018,Verrelli2020}} our screening criteria could not reveal this compound, as a stoichiometric calciated structure, i.e.,  {\ch{Ca_xTiS_2}}, is presently unknown.  However, the robustness of our strategy is validated by the re-discovery of chevrel-{\ch{Mo6S8}} and {\ch{CaMoO3}} phases that have received attention as Ca-electrodes.{\cite{Rogosic2014,Dompablo2016a,Cabello2018}}  Our screening procedure did not identify the following compounds which have been studied computationally before: {\ch{Ca4Fe9O_{17}}} has a low $E_m \sim$720~meV{\cite{Black2020}} but does not have a matching, ordered, charged structure, and {\ch{Ca3Co4O9}} exhibits $E_m \sim 900$~meV{\cite{Torres2019a}} but is significantly unstable (E$^\mathrm{hull} \sim$73~meV/atom{\cite{Jain2013}}). Furthermore, we did not consider {\ch{Ca3Co2O6}} as a possible candidate since previous experimental studies have reported the structure to undergo an irreversible phase transformation upon Ca extraction.{\cite{Torres2019a}}

Using a different computational screening strategy, Zhang et al.{\cite{Zhang2019}} did consider {\ch{CaNb2O4}} as a possible intercalation electrode for Ca batteries. Unfortunately, the authors discarded {\ch{CaNb2O4}} on the basis of the high computed $E_m$ ($\sim$2100 meV) in their work. We believe that Zhang et al.{\cite{Zhang2019}} computed the $E_m$ for Ca migration through the sterically-constrained O--O dumbbell that is shared between adjacent prismatic sites instead of considering the migration mechanism via a quadrilateral planar (see Section~{\ref{sec:revised}} for more detailed explanation) transition state. Ca migration through the O--O dumbbell may have been caused due to one or both of the following reasons: {\emph{i}}) the authors attempted to calculate the full migration path, thereby not having an additional Ca-vacancy that can aid migration (see Figure~{\ref{fig:specbarrier}}b), and {\emph{ii}}) the authors did not initialize their nudged elastic band (NEB{\cite{Henkelman2000}}) calculation via the quadrilateral planar transition state.

Given the scarcity of known good Ca conductors,{\cite{Dompablo2019}} our strategy emphasises the identification of compounds providing facile Ca transport, apart from our criteria including the availability of charged structures and overall thermodynamic stability, while Zhang et al.'s strategy prioritized Ca-intercalation voltages and capacities,{\cite{Zhang2019}} which tends to overlook more practical candidates.

\subsection{Established design rules to identify good ionic conductors}
\label{sec:desing}
It is important to take a look at the three rules to identify good multivalent conductors, initially proposed by Rong et al.\cite{Rong2015}, and examine their applicability with our findings (see Table~{\ref{tbl:discussiontable}} for a summary). Note that the rules were initially devised for the discovery of good Mg candidates and had not been extensively tested in Ca containing compounds. The rules are: 

\begin{enumerate}
\item \emph{Avoid host structure where the active ion, \ch{Ca^{2+}}, resides in its preferred anion coordination environment.} A previous statistical analysis of the ICSD\cite{Bergerhoff1983,Hellenbrandt2004} showed that \ch{Ca^{2+}} highly prefers a coordination number of 8 and above in oxides.\cite{Brown1988} The preferred coordination environment of {\ch{Ca^{2+}}} in chalcogenides, such as S, Se, and Te, has not been yet quantified rigorously. As a result, a good \ch{Ca^{2+}} conductor should display a Ca coordination number $< 8$ in oxides according to this rule. However, one of the two good conductors that we have found in our search, \ch{CaV2O4} does have Ca residing in an 8 coordination environment. In contrast, \ch{CaCu2S2} exhibits a high $E_m$ ($\sim$1622~meV) despite Ca occupying an un-preferred octahedral environment.  With respect to the limited number of structures considered in this study, this rule does not seem to be a good classifier of good vs. bad Ca diffusers. Hence, it is important to update this rule by examining its applicability on an even larger and more diverse dataset.

\item \emph{Reduce changes in the coordination number of the active ion (\ch{Ca^{2+}}) during its migration.} This rule proposes that smaller changes in coordination environment along the migration path of \ch{Ca^{2+}} flattens the potential energy landscape, in turn reducing $E_m$.\cite{Rong2015,Canepa2017a} We could find several exceptions to this rule in the set of materials that we have considered. For example, \ch{CaV2O4}, \ch{CaIrO3}, \ch{CaRh2O4}, \ch{CaMoO3} and \ch{CaV2O5} have similar coordination number change from stable to transition state (i.e., 8$\rightarrow$ 3), yet display widely different migration barriers (654--2072~meV). \ch{CaV3O7}, \ch{Ca2V2O6}, and \ch{CaCu2S2} all have relatively low changes in coordination number, namely 8$\rightarrow$5, 8$\rightarrow$4, and 6$\rightarrow$4, than \ch{CaV2O4}, yet exhibit significantly higher barriers ($>1600$~meV vs.\ 654~meV). Thus, reducing coordination changes alone does not seem to be reducing the $E_m$ in Ca compounds.

\item \emph{Increase the available volume (per number of anions) to \ch{Ca^{2+}} in a given framework.} In ``compact'' frameworks, larger volumes per anion typically correlate with weaker electrostatic interactions of \ch{Ca^{2+}} with the anion (and other cations) and promote a reduction of $E_m$.\cite{Rong2015,Canepa2017a} Generally, larger volumes per anion can be obtained by going from oxides to sulfides (or other chalcogenides) in the same structural framework.\cite{Canepa2017a,Martinolich2019} This rule may not directly apply to non-compact structures, where the active ion resides in tunnel-type (e.g., hollandite \ch{MnO2}) arrangements. Here, we obtain a high barrier of 1622~meV in compact layered-\ch{CaCu2S2} despite a sulfide anion framework. While  the oxide analogue \ch{CaCu2O2} structure has not been reported and we cannot perform a direct barrier comparison with \ch{CaCu2S2}, some of the post-spinel oxides (e.g., \ch{CaV2O4} and \ch{CaRh2O4}) exhibit a lower barrier than \ch{CaCu2S2}, despite being compact frameworks (post-spinel phases are typically stable at high pressures\cite{Hannah2017}).
\end{enumerate}

\noindent In summary, we observe that the above rules may not be applicable to large cation\cite{Shannon1976} migration, e.g., {\ch{K^{+}}} ($\sim$1.64--1.46~\AA{}), {\ch{Rb^{+}}} ($\sim$1.83--1.52~\AA{}), {\ch{Cs^{+}}} ($\sim$1.67--1.88~\AA{}) and {\ch{Ca^{2+}}}  ($\sim$1.34--1.00~\AA{}). In particular, rule No.\ 1 was derived based on observations in compact anion lattices and cannot account for the diversity of geometrical and electrostatic factors appearing in more complex motifs, such as those sampled in our study. For example, it is known that Ca$^{2+}$ does not occupy sites usually occupied by Li$^+$ or Na$^+$ in \ch{LiFePO4} or \ch{NaFePO4},\cite{Kim2020} which is attributed both to the site preference of Ca and the underlying flexibility of the host structure to accommodate Ca.  We attempt to revise the design rules for identifying good conductors, especially for frameworks with large mobile cations, in the following section.

\subsection{Revised design rules for identifying good ionic conductors}
\label{sec:revised}
%
\begin{table*}[ht!]
  \caption{
  Geometric features of the stable and transition states are listed along with the $E_m$ (in meV) for materials considered in this work.  Avg. BL represents the average Ca-anion bond-length in the stable state (in \AA{}). D/A/V stand for diagonal or area or volume at the transition state and in units of \AA{}, \AA{}$^2$, and \AA{}$^3$, respectively. $y_\mathrm{Ca}$ represents the fraction of diagonal/area/volume of the transition state occupied by Ca. Final column indicates any presence of a face-sharing cation at the transition state.
  }
  {\scriptsize
  \label{tbl:discussiontable}
  \begin{tabular*}{\textwidth}{@{\extracolsep{\fill}}lclcllcl@{}}

    \hline \hline
    Material     & $E_m$ &  Stable state & Avg.\ BL & Transition state  &  D/A/V &  $y_\mathrm{Ca}$   & Face-sharing cation?   \\
    \hline \\[-7pt]
    \ch{CaV3O7}  & 2892  & Augmented triangular prism   & 2.44  & Rectangular pyramid    & 7.10$^\mathrm{v}$ &   0.83    &  1 V at 2~\AA{} \\ 
    \ch{Ca2V2O6} & 3235  & Biaugmented triangular prism & 2.51  & Square planar          & 3.36$^\mathrm{a}$ &   0.67    &  N/A \\
    \ch{CaV2O4}  & 654   & Biaugmented triangular prism & 2.45  & Triangular             & 4.56              &   0.86    &  N/A\\ 
    \ch{CaNb2O4} & 785   & Triangular prism             & 2.39  & Rectangular            & 4.42$^\mathrm{a}$ &   0.51    &  N/A \\ 
    \ch{CaMoO3}  & 2072  & Biaugmented triangular prism & 2.55  & Triangular             & 3.82              &   1.03    &  N/A \\  
    \ch{CaIrO3}  & 1219  & Biaugmented triangular prism & 2.46  & Triangular             & 4.32              &   0.91    &  N/A \\ 
    \ch{CaRh2O4} & 1110  & Biaugmented triangular prism & 2.46  & Triangular             & 4.32              &   0.91    &  N/A \\ 
    \ch{CaCu2S2} & 1622  & Octahedron                   & 2.83  & Tetrahedron            & 7.55$^\mathrm{v}$ &   0.78    & 1 Cu at 1.57~\AA{}\\  
    $\alpha$-\ch{CaV2O5} & 1869 & Distorted square antiprism & 2.50 & Triangular         & 3.70              &   1.06    &  N/A \\
    \hline \hline
  \end{tabular*}
    \begin{flushleft}
   $^\mathrm{a}$Shortest diagonal length reported.\\
   $^\mathrm{v}$Volumetric (3D) transition state.
   \end{flushleft}
  }
  \end{table*}

Our motivation to formulate new design rules primarily arises from the following observations among the calculated $E_m$ in this work. Specifically, \emph{i)} Materials with similar structures and Ca coordination environments (at both stable and transition states) exhibit significantly different $E_m$. For example, \ch{CaV2O4} and \ch{CaRh2O4} are both post-spinel structures but show barriers that differ by nearly $\sim$400~meV. \emph{ii)} On the other hand, materials that indeed show similar barriers exhibit remarkably different structures, such as \ch{CaV2O4} and \ch{CaNb2O4}, \ch{CaIrO3} and \ch{CaRh2O4}, and \ch{CaMoO3} and \ch{CaV2O5}. \emph{iii)} The reasons for \ch{CaV3O7} and \ch{Ca2V2O6} to exhibit the highest barriers among the materials considered here, are not immediately obvious.  

We decided to take a deeper look into the geometrical features of the stable and transition states in the structures considered here, and our findings are summarised in Table~{\ref{tbl:discussiontable}}. Note that our analysis is biased by the limited number and diversity of structures that we have sampled and also excludes other thermodynamic (e.g., stability or energy above the convex hull) and electronic (e.g., metallic vs.\ non-metallic) properties that can influence $E_m$. A larger dataset of migration energy profiles across a wider range of structures is required to fully understand and decouple the different contributions to $E_m$, which is beyond the scope of this work.

The biaugmented triangular prism and distorted square antiprism in Table~{\ref{tbl:discussiontable}} correspond to 8-coordinated (stable) sites of Ca and the augmented triangular prism refers to a 7-coordinated site. Avg.\ BL (in \AA{}) refers to the average Ca-anion bond-length in the stable site of each structure. While triangular, rectangular, and square planar are 2D (planar) transition states, rectangular pyramid and tetrahedron are 3D (volumetric). D, A, V in Table~{\ref{tbl:discussiontable}} represent the length of the diagonal (in \AA{}) for a quadrilateral transition state, the area (in \AA{}$^2$) of a triangular transition state, and the volume (in \AA{}$^3$) of a volumetric transition state (i.e., a three-dimensional or a 3D space), respectively. $y_{\mathrm{Ca}}$ represents the fraction of the diagonal/area/volume occupied by Ca at the transition state and the final column of Table~{\ref{tbl:discussiontable}} indicates the presence (or not) of a face-sharing cation and its distance from the migrating Ca, at a volumetric transition state. For identifying the number of Ca-anion bonds at the transition state (and hence to calculate the geometrical features of the transition state), we used the longest Ca-anion bond-length at the stable state as the upper-limit for each structure considered.

Note that all geometrical features reported in Table~{\ref{tbl:discussiontable}} are obtained from a DFT-relaxed geometries of the ground states (endpoint structures). The end-point configurations are typically relaxed before doing NEB and they are not altered during the course of a NEB calculation. The features reported are for a non-migrating Ca ion, i.e., we take the DFT-relaxed endpoint structure and label a Ca-ion that is non-migrating but is symmetrically equivalent to the migrating Ca. Subsequently, we construct a (hypothetical) migration path for the labelled Ca-ion, which mimics the set of images that are used as a starting point for the NEB, locate the transition state along the as-constructed migration path and estimate the geometrical features of the located transition state. Note that we focus on a non-relaxed migration path of a non-migrating Ca-ion to ensure that we include any ``global'' structure relaxations (e.g., any changes in lattice parameters) of the endpoint structures while excluding any ``local'' structure relaxations that result from the relaxation of the end-point configuration at the site of the migrating Ca.

\begin{figure*}[ht!] 
\includegraphics[width=\textwidth]{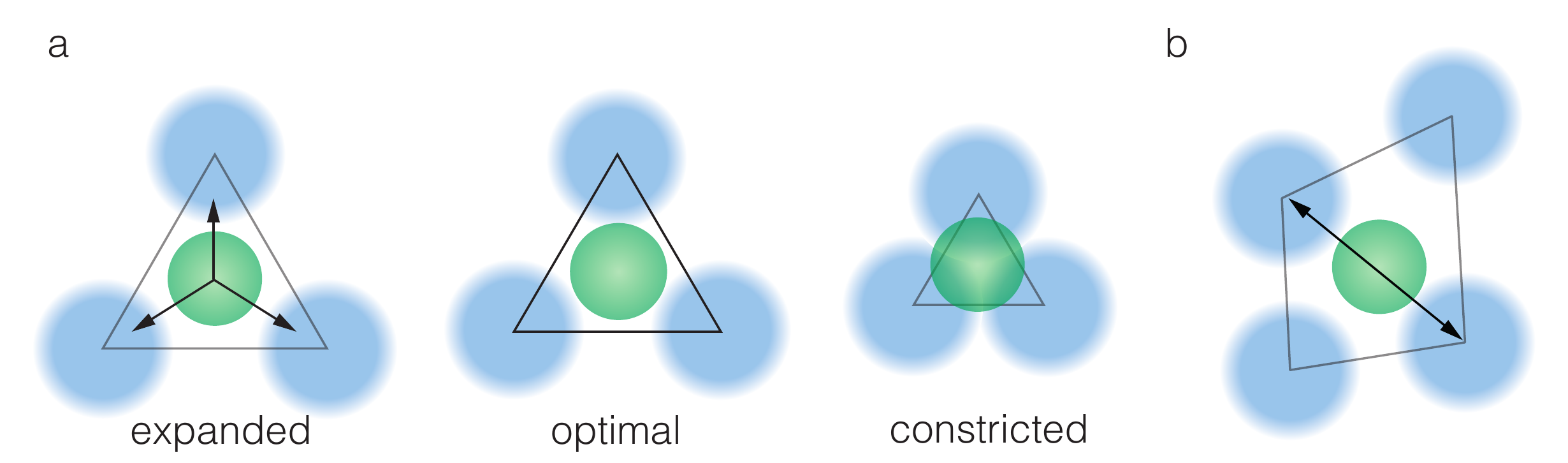}
\caption{
Ca$^{2+}$ ions (green circles) in triangular (panel (a)) and quadrilateral (panel (b)) anion environments, which can constitute possible transition states during migration. Panel (a) classifies three types of triangular environments based on the triangular area, namely expanded, optimal, and constrained. In panel (b) the shortest diagonal of a quadrilateral transition state, along which Ca can share common tangents with anions, is shown.  
}  
\label{fig:discussion} 
\end{figure*}

We can infer from Table~{\ref{tbl:discussiontable}} that the most common stable and transition states are 8-coordinated and 3-coordinates sites, respectively, exhibiting a wide range in $E_m$ from 654-2072~meV. Since the stable 8-coordinated sites do not differ significantly in terms of average BL (2.45--2.51~\AA), we expect the variations in observed $E_m$, in such frameworks, to primarily arise from the differences in the transition state. For example, if Ca is migrating through a triangular transition state, as shown in Figure~{\ref{fig:discussion}}a, an ``optimal'' size of the triangular transition state will ensure a lower $E_m$ compared to scenarios where the triangle is too expanded or too constricted. 

If we assume a rigid sphere model of atoms in a framework, the optimal triangular framework is where the central Ca atom (green spheres in Figure~{\ref{fig:discussion}}a) shares common tangents with the three oxygen atoms (blue spheres) that form the vertices of the triangle. Given an ionic radius\cite{Shannon1976} of Ca$^{2+}$ and O$^{2-}$ to be 1.12~\AA{} and 1.38~\AA{}, respectively, we can calculate the fraction of the triangular area occupied by Ca ($y_\mathrm{Ca}$) under the optimal condition to be $\sim$0.48. In the case of a triangular transition state with S$^{2-}$ vertices (ionic radius of S$^{2-} \sim$1.84~\AA{}),\cite{Shannon1976} optimal $y_\mathrm{Ca}$ will be 0.35. Hence, triangular transition states that have $y_\mathrm{Ca}$ closer to the optimal value will have a lower $E_m$. Similar optimal conditions can be derived for other types of transition states as well. In the case of a quadrilateral-type transition state (Figure~{\ref{fig:discussion}}b), such as, square planar or rectangular, the rigid sphere of Ca$^{2+}$ has to share common tangents with the anion vertices along the shortest diagonal. Thus, for O$^{2-}$ and S$^{2-}$ quadrilateral vertices, the optimal fraction of the diagonal that should be occupied by Ca$^{2+}$ is 0.45 and 0.38, respectively. Analogously, the optimal volume fraction that Ca$^{2+}$ should occupy in a tetrahedron with O$^{2-}$ (S$^{2-}$) vertices is 0.73 (0.44), while in the case of a square pyramid, the optimal volume fraction with O$^{2-}$ vertices is 0.28 ---in a square pyramid, only half of the Ca sphere will be fully contained within the pyramid. The optimal $y_\mathrm{Ca}$ in sulfide frameworks will be lower than oxide analogues due to the larger ionic radius of {\ch{S^{2-}}} compared to {\ch{O^{2-}}}.

We can explain important trends observed in Table~{\ref{tbl:discussiontable}} based on the deviation of the transition state away from the optimal condition. For example, the $y_\mathrm{Ca}$ of the triangular transition state in \ch{CaV2O4} is closer to the optimal triangle by $\sim$0.05 than the $y_\mathrm{Ca}$ of \ch{CaRh2O4}, which causes a lowering of $E_m$ by $\sim$400~meV in \ch{CaV2O4}. Both \ch{CaV2O4} and \ch{CaRh2O4} are post-spinels with nearly-identical stable sites, which indicates the importance of the small deviations in the transition state on the overall barrier. Interestingly, $y_{\mathrm{Ca}}$ in CM- and CF-type post-spinels of composition {\ch{CaMn2O4}} are 0.97 and 0.91, respectively, as calculated from the corresponding DFT-relaxed bulk structures available in the MP. These $y_{\mathrm{Ca}}$ values are in excellent agreement with previously reported{\cite{Dompablo2016}} $E_m$ values, namely $\sim$~1850~meV in CM and $\sim$~1350~meV in CF {\ch{CaMn2O4}}, further validating our optimal $y_\mathrm{Ca}$ criterion.  \ch{CaNb2O4} exhibits a quadrilateral transition state (rectangular, Figure~\ref{fig:specbarrier}b) whose $y_\mathrm{Ca}$ is closer to the optimal value by $\sim$0.16 than the $y_\mathrm{Ca}$ of \ch{Ca2V2O6}, causing a lowering of $\sim$2400~meV of $E_m$ in \ch{CaNb2O4}. Furthermore, \ch{CaRh2O4} and \ch{CaIrO3} both exhibit similar barriers (difference of $\sim$100~meV) owing to similar $y_\mathrm{Ca}$ in the transition state, while the $E_m$ of \ch{CaMoO3} and \ch{CaV2O5} differ marginally ($\sim$200~meV) owing to the change in $y_\mathrm{Ca}$ ($\sim$0.03) at their triangular transition states. Although we do not expect the relationship between the deviation from optimality and $E_m$ to be linear, we can expect a transition state that is closer to its optimality to yield a lower $E_m$. 

The presence (or lack of) non-migrating cations, such as transition metals, or other cations, e.g., S$^{6+}$, P$^{5+}$, etc., that share a ``face'' with a volumetric transition state, can also have a non-negligible impact on the values of $E_m$. Indeed, the high barriers in \ch{CaV3O7} and \ch{CaCu2S2} can be attributed, to a large extent, to the face-sharing V and Cu atoms, respectively, at the transition state during Ca migration. Such face-sharing cations can cause a significant increase in electrostatic potential energy due to electrostatic repulsion, particularly at small separations from the migrating Ca, resulting in a higher barrier. For example, the Ca-Cu distance in \ch{CaCu2S2} at the transition state ($\sim$1.57~\AA{}) is significantly lower than the Ca-S bond distance in the same site ($\sim$2.45~\AA{}), signifying that the Ca-Cu electrostatic repulsion is insufficiently screened by the surrounding anion framework, resulting in the high $E_m$ ($\sim$1622~meV) observed. In general, we expect face-sharing cations to have a higher impact on $E_m$ in oxides than sulfides (or other chalcogenides), in line with previous observations,\cite{Gautam2016} due to the higher ionicity of oxygen-containing bonds and the lack of electrostatic screening by O$^{2-}$ compared to S$^{2-}$,\cite{Canepa2017} partly explaining the higher $E_m$ in \ch{CaV3O7} than \ch{CaCu2S2}. Also, there are examples of electrostatic repulsion forcing multivalent ions to migrate through a sterically-constrained anion-cation-anion dumbbell site instead of a planar or a volumetric transition site,{\cite{Gautam2016}} always resulting in higher barriers. 

Apart from optimal transition states and face-sharing cations, we speculate that structures that exhibit a lower Ca volume fraction change ($\Delta v_\mathrm{Ca}$) during migration, to yield a lower $E_m$. Although we do not have a rigorous set of calculated barriers to justify this suggestion, we highlight the similarities (and differences) with rule No.\ 2 in Sec.~{\ref{sec:desing}}. For a given polyhedral coordination that Ca is occupying, $v_\mathrm{Ca}$ is the fraction of the polyhedral volume occupied by a rigid Ca$^{2+}$ sphere. Therefore, if a migration pathway minimizes $\Delta v_\mathrm{Ca}$ during Ca motion (i.e., reduce the change in $v_\mathrm{Ca}$ between the stable and transition states), it should, in principle, flatten the potential energy landscape, since Ca is moving across polyhedra that are ``similar''. This is analogous  to what is expected by minimising changes in coordination number  during migration, but results in a better metric (both quantitatively and qualitatively) to identify the similarity of polyhedral motifs during ionic migration. However, there is an inherent trade-off associated with maintaining the same $v_\mathrm{Ca}$ and attaining the optimal $y_\mathrm{Ca}$ at transition state, which will eventually decide the magnitude of $E_m$. This trade-off can be approximately considered to be breaking/forming the optimal number of bonds during migration. 

Finally, we suggest that volumetric/planar transition states that have more degrees of freedom (DoF, i.e., number of allowed changes of edge lengths and shapes) to attain better $v_\mathrm{Ca}$-$y_\mathrm{Ca}$ trade-offs, and result in lower $E_m$. Thus, we are hopeful that materials that exhibit a quadrilateral-type transition state, which has more DoF than triangular, and have $E_m$ lower than \ch{CaNb2O4} (shown in this work) will be identified in the near future.

To summarise, we propose the following updated set of design rules for identifying good ionic conductors, especially for structures with large migrating cations:
\begin{enumerate}
\item Find structures that yield close to the optimal diagonal/area/volume (D/A/V) fraction of migrating cation at the transition state (excluding any ``local'' relaxation during migration).
\item In case of volumetric transition states in frameworks, avoid face-sharing cations.
\item Minimise changes in the volume fraction of cation as it migrates across sites (by choosing structures exhibiting transition states with higher degrees of freedom). 
\end{enumerate}

\section{Conclusion}
In summary, we  investigated  the large compositional space  of more than 400 ternary chalcogenides with prototypical formula  {\ch{Ca_iM_jZ_k}}. From this large pool of  materials, 10 materials were studied using first-principles calculations. We found two previously unreported  promising electrode candidates, namely post-spinel \ch{CaV2O4}  and  layered \ch{CaNb2O4}, with Ca migration barriers of $\sim$654~meV and $\sim$785~meV, respectively.  From the analysis of the Ca migration characteristics in the 10 selected compounds, we proposed updated design rules for the identification of good Ca conductors. These rules are to find structures: \emph{i)} with optimal diagonal/area/volume at the transition state, \emph{ii)} without any face-sharing cations at the transition state, and \emph{iii)} that minimise changes in polyhedral volumes during ion migration. While our study centered on Ca-based materials for positive electrodes, our findings may be extended to other compounds hosting large mobile cations.

\section{Methodology}
\label{sec:methods}
\subsection{\emph{Ab initio} Thermodynamics of Conversion and Intercalation Voltages}
An intercalation battery based on the hypothetical \ch{Ca_nTM_jZ_k} cathode electrode implies the reversible extraction/insertion of \ch{Ca^{2+}} ions from/into the \ch{Ca_nTM_jZ_k} framework following the spontaneous reaction of Eq.~\ref{eq:volatgereaction}.
\begin{equation}
\mathrm{TM_jZ_k} + \mathrm{nCa} \xrightarrow{-\Delta G_\mathrm{inter.}} \mathrm{Ca_nTM_jZ_k}
\label{eq:volatgereaction}
\end{equation}
where \ch{n} sets the \ch{Ca} concentration and $-\Delta G_\mathrm{inter.}$ is the change of Gibbs energy at 0~K, where  the DFT total energies are approximated as the Gibbs energies (i.e., $E \approx G$). By doing so, we neglect the $pV$ and entropic contributions to the Gibbs energy. From Eq.~\ref{eq:volatgereaction} we estimate the average intercalation voltage, $V_{inter.}$, as expressed in Eq.~\ref{eq:voltage}.
\begin{equation}
V_\mathrm{inter.} = -\frac{\Delta G_\mathrm{inter.}}{2nF} = -\frac{E_\mathrm{Ca_nTM_jZ_k} - \left[E_\mathrm{TM_jZ_k} + n\mu_\mathrm{Ca}\right]}{2nF}
\label{eq:voltage}
\end{equation}
where $\mu_\mathrm{Ca}$ is the Ca chemical potential (set to that of  Ca metal) and $F$ is the Faraday constant. The value 2 in the denominator of Eq.~\ref{eq:voltage} takes into account the two electrons released upon oxidation of \ch{Ca} to \ch{Ca^{2+}}. Note that the reaction represented by Eq.~\ref{eq:volatgereaction} can be either topotactic or non-topotactic in nature. Computed intercalation voltages are reported in Table~\ref{tbl:itercalationdata} and Figure~\ref{fig:voltages}.

An alternative process to Ca intercalation is the conversion of a hypothetical \ch{TM_jZ_k} cathode upon reduction with Ca, leading typically to the irreversible formation of stable binary/ternary oxides (e.g., CaO and TMO) that can detrimentally stop any electrochemical activity. Previous studies have examined the competition between intercalation and conversion reactions in a large dataset of multivalent TM oxide, sulfide, and selenide chemistries.\cite{Canepa2017,Hannah2018} As an example, a possible conversion reaction is shown in Eq.~\ref{eq:conversionreaction} for a hypothetical \ch{TM_jZ_k} forming CaZ and \ch{TM_jZ_{k-n}}.
\begin{equation}
\mathrm{\mathrm{TM_jZ_k}} + n\mathrm{Ca} \xrightarrow{-\Delta G_\mathrm{conv.}} n\mathrm{CaZ} + \mathrm{TM_jZ_{k-n}}
\label{eq:conversionreaction}
\end{equation}
A conversion voltage can be linked to such reactions, as represented by Eq.~\ref{eq:conv_voltage}.
\begin{equation}
V_\mathrm{conv.} = -\frac{\Delta G_\mathrm{conv.}}{2nF} = -\frac{nE_\mathrm{CaZ} + E_{\mathrm{TM_jZ_{k-n}}} - \left[E_\mathrm{TM_jZ_k} + n\mu_\mathrm{Ca}\right]}{2nF}
\label{eq:conv_voltage}
\end{equation}
For intercalation reactions to be thermodynamically favourable, $V_\mathrm{inter.}$ needs to be higher than $V_\mathrm{conv.}$.  Another way to assess whether the intercalation of Ca is favoured over undesired conversion reactions is to examine the thermodynamic stability of the discharged \ch{Ca_iTM_jZ_k} compound with respect to a combination of all possible compounds that can exist at the same composition, which is captured by the E$^\mathrm{hull}$.  A potential cathode that exhibits E$^\mathrm{hull}$~=~0~meV/atom will strongly favour intercalation reactions over conversion upon reduction with Ca. 

\subsection{First-principles Calculations}

To assess the suitability of  the \ch{Ca_iTM_jZ_k}-type compounds  for Ca electrodes, we computed the migration energy $\mathrm{E}_m$ using the NEB method\cite{Henkelman2000,Henkelman2000a} together with DFT calculations.\cite{hohenberg_inhomogeneous_1964,Kohn1965}   In DFT, the total energy was approximated by the  generalized
gradient approximation (GGA) exchange-correlation functional as parameterized by Perdew, Burke, and Ernzerhof (or PBE),\cite{Perdew1996}  which was evaluated using the code VASP. \cite{kresse_efficient_1996,kresse_efficiency_1996} The electronic wavefunctions were
expanded as plane-waves up to a kinetic
energy cut-off of 520 eV and combined with the projector-augmented-wave
(PAW) potentials for the core
electrons.\cite{kresse_ultrasoft_1999} The PAW potentials used were, Ca\_sv (06Sep2000 $\mathrm{3s3p4s}$),  Cu (05Jan2001 \ch{d^{10}p^{1}}), Ir (06Sep2000 \ch{s^1d^8}), Mo\_pv (08Apr2002 $\mathrm{4p5s4d}$), Nb\_pv (08Apr2002 $\mathrm{4p5s4d}$), O (08Apr2002 $\mathrm{2s2p}$), Rh (06Sep2000 \ch{s^1d^8}),  S (17Jan2003 $\mathrm{s^2p^4}$), and V\_pv (07Sep2000 \ch{p^6d^4s^1}). The total energies in DFT were integrated using a  $k$-point density of at least 1000/atom, and
converged within 10$^{-6}$~eV/cell, without imposing symmetry restrictions. The
atomic forces and stresses were converged to within 10$^{-2}$ eV/\AA{} and 0.29 GPa, respectively. 

For the estimations of the Ca intercalation voltages in \ch{Ca_iTM_jZ_k} compounds,  which often behave as highly-correlated systems,\cite{Wang2006,Zhou2006,Franchini2007,Hinuma2008,Jain2011,SaiGautam2018} we employed the rotationally invariant Hubbard $U$ approach to GGA, as proposed by Dudarev \emph{et al.}\cite{Dudarev1998}, resulting in a GGA+$U$ (or PBE+$U$) Hamiltonian. GGA+$U$  corrects for the self-interaction error introduced by the highly localized $d$ orbitals of the redox species V, Cu, Nb, and Mo. Note that the empirical GGA+\emph{U} correction has been extensively calibrated to compute intercalation voltages in battery materials.\cite{Wang2006,Jain2011,SaiGautam2018} The specific $U$ values used were 3.10~eV for V, 4.00~eV for Cu, 1.50~eV for Nb, and 4.38~eV for Mo, as previously reported by Jain et al.\cite{Jain2011}

Following the prescription by Chen \emph{et al.},\cite{Chen2019} we implemented model supercells that always introduced a minimum distance $\geq$8~\AA{} between the Ca ions participating in the ion migration, to minimize the fictitious interactions between periodic images, in our NEB{\cite{Henkelman2000,Henkelman2000a,Sheppard2008}} calculations. We fully relaxed (\emph{i.e.}, coordinate, volume and shape) the initial and final states defining the Ca$^{2+}$ migration paths in our NEB until the forces on the atoms were lower than 10$^{-2}$ eV/\AA{}$^{-1}$.  Note that we used the GGA functional in our initial/final state and the overall NEB calculation as well, since the GGA has been shown to be more appropriate than GGA+$U$ to predict ion migration paths.\cite{Rong2015} The elastic bands in all materials considered is constructed using seven distinct, equally-spaced images, between the ground-state structures (i.e., the endpoints). The forces on the elastic band are converged to within 0.05 eV \AA{}$^{-1}$.  All $E_m$ estimations were done in the regime of dilute vacancy, \emph{i.e.}, one vacancy is created in the Ca lattice. Notably, the introduction of Ca-vacancies in the potential electrodes (as required for NEB calculations) were compensated by the spontaneous oxidation of nearby transition metal(s).

\begin{acknowledgement}  
P.C. acknowledges funding from the National Research Foundation under his NRF Fellowship
NRFF12-2020-0012 and support from the Singapore Ministry of Education Academic Fund Tier 1 (R-284-000-186-133).  G.S.G.\ acknowledges financial support from the Indian Space Research
Organization-Space Technology Cell at the Indian Institute of Science, under project code number ISTC/MET/SGG/451. 
The computational work was performed on resources of the National
Supercomputing Centre, Singapore (\url{https://www.nscc.sg}).
\end{acknowledgement}

\begin{tocentry} 

\includegraphics{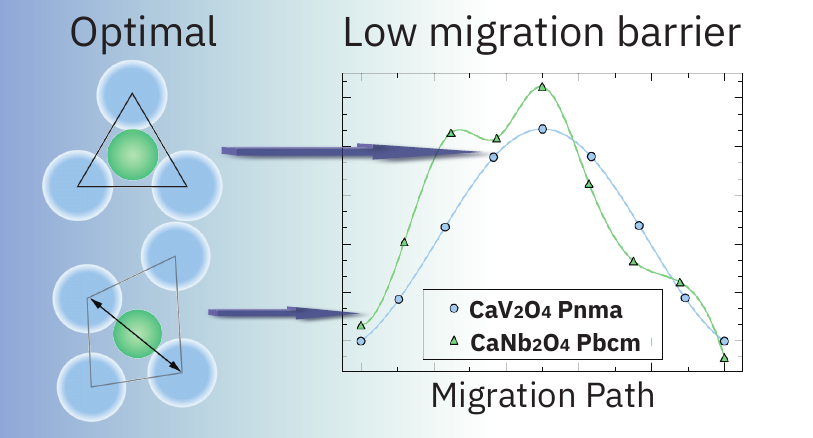}

\end{tocentry}

\begin{suppinfo} 
The supporting information contains: \emph{i}) The structural and thermodynamic stabilities of 412 ternary chalcogenides, \emph{ii}) A model showing the maximum tolerable migration barrier vs.\ particle size, rate, and temperature, and  \emph{iii}) The computed migration energy profiles of ternary chalcogenides considered. 
\end{suppinfo}

\bibliography{biblio}

\end{document}